\documentclass[eqsecnum,10pt,a4paper]{article}
\usepackage{amsmath}
\usepackage{graphicx}
\numberwithin{equation}{section}

\begin{document}

\title{Partition Function of the Schwarzschild Black Hole}

\author{Jarmo M\"akel\"a\footnote{Vaasa University of Applied Sciences,
Wolffintie 30, 65200 Vaasa, Finland, email: jarmo.makela@puv.fi}}

\maketitle

\begin{abstract} 

We consider a microscopic model of a stretched horizon of the Schwarz\-schild black hole. In our model the stretched
horizon consists of a finite number of discrete constituents. Assuming that the quantum states of the Schwarzschild
black hole are encoded in the quantum states of the constituents of its stretched horizon in a certain manner we 
obtain an explicit, analytic expression for the partition function of the hole. Our partition function predicts, 
among other things, the Hawking effect, and provides it with a microscopic, statistical interpretation.

\end{abstract}

{\footnotesize{\bf PACS}: 04.60.Nc, 04.70.Dy}

{\footnotesize{\bf Keywords}: Constituents of spacetime, partition function, Hawking effect.}

\section{Introduction}

During the course of time the problem of how to quantize gravity has
been approached in various ways by a number of authors. Despite the
great diversity in their opinions, there is at least one point where
all serious physicists specializing in quantum gravity will agree:
whatever the forthcoming generally accepted quantum theory of
gravitation may be, it should be able to provide a statistical,
microscopic interpretation to the so-called {\it Hawking effect}.
According to this effect black holes are not completely black, but
they emit radiation with a characteristic temperature, which for a
Schwarzschild black hole with Schwarzschild mass $M$ is given by the
Hawking temperature
\begin{equation}\label{eq:1.1}
T_H := \frac{1}{8\pi}\frac{\hbar c^3}{k_B G}\frac{1}{M}.
\end{equation}
Even though the Hawking effect was discovered already about 35 years
ago as a straightforward consequence of the quantum theory of fields
in curved spacetime \cite{yksi}, we may safely say that nobody
really understands it. Why does a black hole radiate? Is the
radiation of a black hole a somewhat similar process as is the
radiation of ordinary matter, where the atoms of the matter perform
jumps between different quantum states and, as a consequence,
photons are emitted? If so, what are the ``atoms'' of spacetime
which are supposed to constitute, among other things, black holes? What are their
quantum states? Obviously, questions of this kind take us to the
very foundations of the concepts of space and time. What is the
proper microscopic description of space and time?

When considering these questions we need a microscopic model of a
black hole which, even if not necessarily correct, at least allows
us to address our questions in precise terms. We are going to
construct such a model in this paper for the Schwarzschild black
hole. Whenever we go over from a microscopic, statistical
description of any system to its macroscopic, thermodynamical
description, the key role is played by the {\it partition function}
\begin{equation}\label{eq:1.2}
Z(\beta) := \sum_n g(E_n)e^{-\beta E_n}
\end{equation}
of the system. In Eq. (1.2) $n$ labels the possible different total
energies $E_n$ of the system, $\beta$ is the inverse temperature of
the system, and $g(E_n)$ is the number of states associated with the
same total energy $E_n$. We construct a microscopic model of what we
call as a ``stretched horizon'' of a Schwarzschild black hole, and
we write the partition function of the Schwarzschild black hole from
the point of view of an observer on its stretched horizon. Using our
partition function we obtain, among other things, the Hawking
effect, and provide it with a statistical, microscopic
interpretation.

We begin our considerations in Section 2 with a definition of the
concept of stretched horizon of the Schwarzschild black hole. In
broad terms, our stretched horizon may be described as a space-like
two-sphere just outside of the event horizon of the hole. An
observer at rest with respect to the stretched horizon has a certain
proper acceleration, and we require that when the Schwarzschild mass
of the hole is changed, its stretched horizon will also change, but
in such a way that the proper acceleration of the observer stays
unchanged. In other words, we require that no matter what may happen
to the black hole, an observer on the stretched horizon will always
feel the one and the same proper acceleration. One finds, quite
remarkably, that if such an observer is initially close to the event
horizon of the black hole, he will stay close to the event horizon
even when the Schwarzschild mass of the hole is changed. In this
respect our notion of stretched horizon really makes sense.

The partition function $Z(\beta)$ of Eq. (1.2) involves the concept
of energy. Unfortunately, the concept of energy is very problematic
in general relativity. Nevertheless, it turns out possible to
define, beginning from the so-called Brown-York energy
\cite{kaksi}, the notion of energy of the Schwarzschild black hole
from the point of view of an observer on its stretched horizon. The
resulting expression for the energy turns out to be, in SI units,
\begin{equation}\label{eq:1.3}
E = \frac{ac^2}{8\pi G}A,
\end{equation}
where $a$ is the proper acceleration of an observer on the stretched
horizon, and $A$ is the area of the horizon. Since the proper
acceleration $a$ is assumed to be a constant, the energy of the hole
depends, from the point of view of an observer on the stretched
horizon, on the area $A$ of the horizon only.

To consider the microscopic origin of the Hawking effect we need an
appropriate microscopic model of the stretched horizon. Simple
models are often the best models, and in this paper we shall settle
in a model that  can hardly be challenged in simplicity. We shall
assume that the stretched horizon consists of a finite number of
discrete constituents, each of them contributing to the stretched
horizon an area, which is a non-negative integer times a constant.
More precisely, we write the area of the stretched horizon, in SI
units, as:
\begin{equation}\label{eq:1.4}
A = \alpha \ell_{Pl}^2(n_1 + n_2 + ... + n_N).
\end{equation}
In this equation $N$ is the number of the constituents, $n_1, n_2,
..., n_N$ are non-negative integers determining their quantum
states, $\alpha$ is a number to be determined later, and
\begin{equation}\label{eq:1.5}
\ell_{Pl} := \sqrt{\frac{\hbar G}{c^3}} \approx 1.6\times 10^{-35}m
\end{equation}
is the Planck length. We shall not specify what the constituents of
the stretched horizon actually are. Since each constituent
contributes a certain area to the stretched horizon, we shall call
the quantum states of the constituents determined by the quantum
numbers $n_1, n_2, ..., n_N$ as their {\it area
eigenstates}~\cite{kolmea}.

After introducing our model in Section 2 we shall proceed, in
Section 3, to the calculation of the partition function of the
Schwarzschild black hole. Following the commonly accepted---although
still unproved---wisdom that the quantum states of a black hole are
somehow encoded in its horizon [4-6], we
shall assume that for each stationary quantum state of a black hole
there exists a unique quantum state, determined by the quantum
numbers $n_1, n_2,..., n_N$, of its stretched horizon. The
calculation of the partition function is based on what we shall call
as a {\it statistical postulate} of our model. According to this
postulate the microscopic states of the black hole are identified
with the combinations of the non-vacuum area eigenstates of the
constituents of its stretched horizon. Eqs. (1.3) and (1.4) imply
that the possible energies of the black hole are, from the point of
view of an observer on its stretched horizon, of the form:
\begin{equation}\label{eq:1.6}
E_n = n\alpha\frac{\hbar a}{8\pi c},
\end{equation}
where
\begin{equation}\label{eq:1.7}
n = n_1 + n_2 + ... + n_N.
\end{equation}
Hence it follows from the statistical postulate that the number of
microscopic states associated with the same energy $E_n$ is the same
as is the number of ways of expressing a given positive integer $n$
as a sum of at most $N$ positive integers $n_1, n_2, ..., n_N$. More
precisely, it is the number of ordered strings $(n_1,n_2,...,n_m)$,
where $1 \le m \le N$, $n_j \in \lbrace 1, 2,...\rbrace$ for all $j
= 1, 2,..., m$ and $n_1 + n_2 +...+ n_m = n$. This number, which
depends on $n$ and $N$ only, gives the function $g(E_n)$ in
Eq. (1.2), and it may be calculated explicitly. It is most gratifying
that the resulting partition function $Z(\beta)$ of the
Schwarzschild black hole may be also calculated explicitly, yielding
a surprisingly simple expression:
\begin{equation}\label{eq:1.8}
Z(\beta) = \frac{1}{2^{\beta T_C} - 2}\left\lbrack 1 -
\left(\frac{1}{2^{\beta T_C} - 1}\right)^{N+1}\right\rbrack,
\end{equation}
where we shall call the temperature
\begin{equation}\label{eq:1.9}
T_C := \frac{\alpha\hbar a}{8(\ln 2)\pi k_Bc}
\end{equation}
as the {\it characteristic temperature} of the hole.

All thermodynamical properties of the Schwarzschild black hole will
follow, in our model, from the partition function of Eq. (1.8). In
Section 4 we shall consider the dependence of the energy of the hole
on its absolute temperature $T$. The most important outcome of those
considerations is a result, which will be investigated in details in
Section 5, that when $T = T_C$, the Schwarzschild black hole
performs a {\it phase transition}, where the constituents of its
stretched horizon jump, in average, from the vacuum to the second
excited states. More precisely we shall see, in the large $N$ limit,
that when $T < T_C$, all constituents of the stretched horizon,
except one, are in vacuum, whereas when $T$ is slightly higher than
$T_C$, the constituents are, in average, in the second excited
states. Since the constituents are in vacuum, when $T < T_C$, there
is no black hole with a temperature less than its characteristic
temperature $T_C$, and in this sense the characteristic temperature
$T_C$ is the lowest possible temperature a Schwarzschild black hole
may have. Choosing
\begin{equation}\label{eq:1.10}
\alpha = 4\ln 2
\end{equation}
in Eq. (1.9) one finds in the leading approximation, when using the
natural units, where $\hbar = c = G = k_B = 1$:
\begin{equation}\label{eq:1.11}
T_C = \frac{a}{2\pi} = \left(1 -
\frac{2M}{r}\right)^{-1/2}\frac{1}{8\pi M}.
\end{equation}
This is the lowest possible temperature measured by an observer on
the stretched horizon of the Schwarzschild black hole with
Schwarzschild mass $M$, and $r$ is the radial Schwarzschild
coordinate of that observer. In our model one may interpret the
effective, non-zero temperature of the black hole as an outcome of a
thermal radiation emitted by the hole, when the constituents of its
stretched horizon perform transitions from the excited states to the
vacuum. The factor $(1-\frac{2M}{r})^{-1/2}$ is the blue shift
factor of the temperature, and using the Tolman relation
\cite{nelja} one finds that when the possible backscattering effects
of the radiation from the spacetime geometry are neglected, the
temperature measured by an observer at a faraway infinity for the
radiation emitted by the black hole is, in SI units,
\begin{equation}\label{eq:1.12}
T = \frac{1}{8\pi}\frac{\hbar c^3}{k_BG}\frac{1}{M},
\end{equation}
which is exactly the Hawking temperature $T_H$ of Eq. (1.1). Hence we
may really obtain the Hawking effect from our model, and provide it
with a microscopic, statistical interpretation.

Closely related to the Hawking effect is the so-called {\it
Bekenstein-Hawking entropy law}, which states that black hole
possesses entropy which, in the natural units, is one-quarter of its
event horizon area $A_H$ or, in SI units \cite{viisi},
\begin{equation}\label{eq:1.13}
S_{BH} = \frac{1}{4}\frac{k_Bc^3}{\hbar G}A_H.
\end{equation}
In Section 6 we shall consider, beginning from the partition
function $Z(\beta)$ of Eq. (1.8), the entropic properties of the
Schwarzschild black hole. Since the stretched horizon lies just
outside of the event horizon of the Schwarzschild black hole, we may
equate, for all practical purposes, the stretched horizon area $A$
of the hole with its even horizon area $A_H$. We shall show in
Section 6 that when $T = T_C$, the Bekenstein-Hawking entropy law
of Eq. (1.13) is exactly reproduced, except that $A_H$ has been
replaced with $A$. However, if $T > T_C$, the situation is somewhat
different. When $T$ is slightly greater than $T_C$, the black hole
has just performed a phase transition, where the constituents of its
stretched horizon have jumped from the vacuum to the second excited
states, and the stretched horizon possesses a {\it critical area}
\begin{equation}\label{eq:1.14}
A_{crit} := 8N\ell_{Pl}^2\ln 2,
\end{equation}
which has been obtained from Eq. (1.4) by putting $\alpha = 4\ln 2$
and $n_1 = n_2 =...= n_N = 2$. If $T > T_C$ then $A \ge A_{crit}$,
and one may obtain for the black hole entropy an expression:
\begin{equation}\label{eq:1.15}
S = \frac{1}{4\ln 2}\frac{k_Bc^3}{\hbar G}A\ln\left(\frac{2A}{2A -
A_{crit}}\right) + Nk_B\ln\left(\frac{2A -
A_{crit}}{A_{crit}}\right).
\end{equation}
This expression provides a modification, involving logarithmic
terms, of the Bekenstein-Hawking entropy law. As one may observe,
the Bekenstein-Hawking entropy law of Eq. (1.13) is exactly
reproduced, when $A = A_{crit}$.

We shall close our discussion in Section 7 with some concluding
remarks. Unless otherwise stated, we shall always use the natural
units, where $\hbar = c = G = k_B = 1$.

\section{The Model}

\subsection{Stretched Horizon}

In the presence of a Schwarzschild black hole the spacetime geometry
is described by the Schwarzschild metric
\begin{equation}\label{eq:2.1}
ds^2 = -(1 - \frac{2M}{r})\,dt^2 + \frac{dr^2}{1 - \frac{2M}{r}} +
r^2\,d\theta^2 + r^2\sin^2\theta\,d\phi^2,
\end{equation}
where $M$ is the Schwarzschild mass of the hole, and we have used
the Schwarzschild coordinates, together with the natural system of
units. The Schwarzschild black hole has the so-called event horizon,
where $r = 2M$. There are good reasons to believe that the
thermodynamical properties of a black hole may ultimately be reduced
to the properties of its event horizon, and hence it becomes
necessary to construct an appropriate microscopic model of the
horizon. Unfortunately, the Schwarzschild coordinates are very
ill-behaving at the horizon, although they are very simple
elsewhere. Of course, there are several system of coordinates, which
behave well at the horizon, but all of them have a disadvantage of
being pretty complicated. Because of these difficulties our idea is
actually to construct a model not of the event horizon itself, but
rather of a two-dimensional space, which lies just outside of the
horizon. More precisely, we shall consider a space-like, closed
two-sphere, where both $r$ and $t$ are constants such that $r > 2M$.
Taking $r$ closer and closer to $2M$ we may investigate the
properties of the event horizon by means of those of our two-sphere.
With a slight misuse of terminology we shall call our two-sphere,
for the sake of brevity and simplicity, as a {\it stretched
horizon}.

On the stretched horizon, where both $r$ and $t$ are constants in
Schwarzschild spacetime, the proper acceleration vector field is
\begin{equation}\label{eq:2.2}
a^\mu := u^\alpha u^\mu_{;\alpha},
\end{equation}
where the semicolon denotes the covariant derivative, and $u^\mu$ is
the normed, future directed tangent vector field of the congruence
of the world lines of those points of spacetime, where all of the
coordinates $r$, $\theta$ and $\phi$ are constants. The only
non-zero component of the vector field $u^\mu$ is
\begin{equation}\label{eq:2.3}
u^t = \left(1 - \frac{2M}{r}\right)^{-1/2},
\end{equation}
and the only non-zero component of the proper acceleration vector
field $a^\mu$ is
\begin{equation}\label{eq:2.4}
a^r = u^t u^r_{;t} = \Gamma^r_{tt}u^t u^t = \frac{M}{r^2}.
\end{equation}
The vector field $a^\mu$ is spacelike, and it is orthogonal both to
the vector field $u^\mu$ and to the stretched horizon. The norm of
the vector field $a^\mu$ is
\begin{equation}\label{eq:2.5}
a := \vert\vert a^\mu\vert\vert = \sqrt{a_\mu a^\mu} = \left(1 -
\frac{2M}{r}\right)^{-1/2}\frac{M}{r^2},
\end{equation}
and it gives the acceleration of particles in a radial free fall
towards the black hole from the point of view of an observer at rest
with respect to the coordinates $r$, $\theta$ and $\phi$. One finds
that the proper acceleration $a$ is otherwise similar to the
gravitational acceleration $\frac{M}{r^2}$ predicted by Newton's
theory of gravitation, except that we have corrected the Newtonian
acceleration by the ``blue shift factor'' $(1 -
\frac{2M}{r})^{-1/2}$. At the event horizon, where $r = 2M$, the
proper acceleration $a$ becomes infinite. However, a faraway
observer measures for particles in a free fall at the horizon a
finite proper acceleration
\begin{equation}\label{eq:2.6}
\kappa := \frac{1}{4M},
\end{equation}
which has been obtained from the proper acceleration $a$ of Eq. (2.5)
such that we have neglected the blue shift factor, and replaced $r$
by $2M$. The quantity $\kappa$ is known as the {\it surface gravity}
of the Schwarzschild black hole \cite{kuusi}, and in SI units it
takes the form:
\begin{equation}\label{eq:2.7}
\kappa = \frac{c^4}{4GM}.
\end{equation}

The properties of the stretched horizon may be varied by changing
the values taken by $r$ and $M$. However, whenever we change the
values of $r$ and $M$, we do that in such a way that the proper
acceleration $a$ on our surface is kept as a constant. This means
that if $r$ and $M$ take on infinitesimal changes $dr$ and $dM$,
then
\begin{equation}\label{eq:2.8}
da = \frac{\partial a}{\partial M}\,dM + \frac{\partial a}{\partial
r}\,dr = 0.
\end{equation}
Using Eq. (2.5) one finds that this requirement implies the following
relationship between $r$ and $M$:
\begin{equation}\label{eq:2.9}
\frac{dr}{dM} = \frac{r}{M}\frac{r-M}{2r-3M}.
\end{equation}
When our surface approaches the horizon, we have:
\begin{equation}\label{eq:2.10}
\lim_{r\rightarrow 2M^+}(\frac{dr}{dM}) = 2,
\end{equation}
and hence $dr \approx 2dM$ in the immediate vicinity of the horizon.
Since $r = 2M$ at the horizon, this result implies that the
stretched horizon stays close to the event horizon even when the
Schwarzschild mass $M$ of the hole is changed.


Among the first researchers to introduce the notion of stretched
horizon in black hole physics were Susskind, Thorlacius and Uglum in
\cite{nelja1}. However, the stretched horizon introduced by
Susskind~{\it et~al.} is essentially different from ours in the
sense that whereas we keep the proper acceleration $a$ on the
stretched horizon as a constant, Susskind {\it et al.} keep as a
constant the difference between the area of the stretched horizon
and that of the event horizon. More precisely, they write the
stretched horizon area $A$~as:
\[
A = A_H + \delta,
\]
where $A_H$ is the event horizon area of the black hole under
consideration, and $\delta$ is a positive constant. For a
Schwarzschild black hole this condition implies that between the
infinitesimal changes $dr$ and $dM$ in the Schwarzschild coordinate
$r$ and the Schwarzschild mass $M$ there is a relationship:
\[
\frac{dr}{dM} = \frac{4M}{r},
\]
which is totally different from Eq. (2.9). The main reason for taking
the proper acceleration $a$ on our stretched horizon to be a
constant is that it simplifies the calculation of the partition
function of the Schwarzschild black hole. Ultimately, our aim is to
obtain the Hawking effect and the other well known thermodynamical
properties of the Schwarzschild black hole from the point of view of
an observer in a spacelike infinity. Our strategy is to consider
first the thermodynamical properties of the Schwarzschild black hole
from the point of view of an observer on the stretched horizon,
where $a = constant$, and then find what the results mean in the
rest frame of a faraway observer. We shall see later that one of the
advantages of using the stretched horizons, where $a = constant$,
instead of using, say, the stretched horizons of Susskind {\it et
al.}, is that from the point of view of an observer on our stretched
horizon the temperature of the radiation emitted by a black hole
remains the same during its evaporation, and there is not any
drastic temperature increase at the final stages of the evaporation.

\subsection{Energy}

As it is well known, the Schwarzschild mass $M$ of the Schwarzschild
black hole may be written in terms of its surface gravity $\kappa$
and event horizon area $A$ as \cite{kuusi}
\begin{equation}\label{eq:2.11}
M = \frac{\kappa}{4\pi}A.
\end{equation}
Indeed, if one substitutes $1/4M$ for $\kappa$, and $A = 4\pi(2M)^2
= 16\pi M^2$ for the event horizon area on the right hand side of
Eq. (2.11), one recovers the Schwarzschild mass $M$. The
Schwarzschild mass $M$ may be regarded, in the natural units, as the
energy of the Schwarzschild black hole from the point of view of a
distant observer at rest with respect to the hole.

In general, the concept of energy is very problematic in general
relativity. This is pretty harmful, because the concept of energy
holds the central stage in thermodynamics. One of the suggestions
given during the course of time for the concept of energy in general
relativity is the so-called {\it Brown-York energy} \cite{kaksi}
\begin{equation}\label{eq:2.12}
E_{BY} := -\frac{1}{8\pi}\oint(k - k_0)\,dA,
\end{equation}
where $k$ is the trace of the exterior curvature tensor on a closed
spacelike two-surface embedded in a spacelike hypersurface, where
the time coordinate $t = constant$, and $k_0$ is the trace of the
exterior curvature tensor, when the two-surface has been embedded in
flat spacetime. $dA$ is the area element on the two-surface, and we
have integrated over the whole two-surface. In Schwarzschild
spacetime the Brown-York energy takes, when the closed two-surface
under consideration is a two-sphere, where $r = constant$, the form
\cite{kaksi}:
\begin{equation}\label{eq:2.13}
E_{BY} = r\left(1 - \sqrt{1 - \frac{2M}{r}}\right).
\end{equation}

The Brown-York energy may be viewed, in some sense, as the energy
of the gravitational field inside a closed, space-like two-surface
of spacetime. The general expression of Eq. (2.12) for the
Brown-York energy $E_{BY}$ may be justified by means of an analysis
of the Hamilton-Jacobi formulation of classical general relativity.
It is interesting that in Schwarzschild spacetime one may obtain the
expression Eq. (2.13) for $E_{BY}$ without any reference to the
Hamilton-Jacobi formulation of general relativity. Instead, one
considers the mass-energy flown through the two-sphere $r =
constant$ during the formation of the Schwarzschild black hole by
means of the gravitational collapse.

To begin with, suppose that a particle falls through the two-sphere
$r = constant$ such that its energy on the two-surface from the
point of view of a distant observer is $\epsilon$. From the point of
view of an observer on the two-surface the energy of the particle,
however, is not $\epsilon$, but
\begin{equation}\label{eq:2.14}
{\tilde\epsilon}(r) =\left(1 - \frac{2M}{r}\right)^{-1/2}\epsilon.
\end{equation}
In other words, we must blue shift the energy of the particle by the
factor $(1 - \frac{2M}{r})^{-1/2}$. Hence it follows that if the
Schwarzschild mass of the Schwarzschild black hole has been
increased by $dM$, the increase in its mass from the point of view
of an observer on the two-sphere $r = constant$ is
\begin{equation}\label{eq:2.15}
dm(r) = \left(1 - \frac{2M}{r}\right)^{-1/2}\,dM.
\end{equation}
The total mass $m(r)$ of the hole is, from the point of view of an
observer on the two-sphere, the mass of the matter, which has fallen
through the two-sphere during the formation of the black hole, and
it is given by
\begin{equation}\label{eq:2.16}
m(r) = \int_0^M \left(1 - \frac{2M'}{r}\right)^{-1/2}\,dM' =
r\left(1 - \sqrt{1 - \frac{2M}{r}}\right),
\end{equation}
which is exactly the Brown-York energy of Eq. (2.13).

A similar reasoning may be applied when we attempt to find an
expression for the energy of the Schwarzschild black hole from the
point of view of an observer on its stretched horizon. Using
Eqs. (2.5) and (2.15) we find that the increase $dm(r)$ in the
mass of the hole from the point of view of our observer may be
written in terms of the proper acceleration $a$ as:
\begin{equation}\label{eq:2.17}
dm(r) = a\frac{r^2}{M}\,dM.
\end{equation}
During the formation of the black hole we keep the proper
acceleration $a$ on the stretched horizon as a constant. As a
consequence, $r$ is not a constant, when the mass of the hole is
increased, but between the infinitesimal changes $dr$ and $dM$ of
$r$ and $M$ there is the relationship (2.9). Hence we find that
the infinitesimal change $dm(r)$ may be expressed in terms of the
infinitesimal change $dr$ as:
\begin{equation}\label{eq:2.18}
dm(r) = a\frac{r^2}{M}\frac{dM}{dr}\,dr = a\frac{2r-3M}{r-M}r\,dr.
\end{equation}
The increase $dr$ in the radial coordinate $r$ results an increase
\begin{equation}\label{eq:2.19}
dA = d(4\pi r^2) = 8\pi r\,dr
\end{equation}
in the area of the two-sphere $r = constant$, and hence we may write
Eq. (2.18) as:
\begin{equation}\label{eq:2.20}
dm(r) = \frac{1}{8\pi}a\frac{2r-3M}{r-M}\,dA.
\end{equation}
Just outside of the event horizon we must consider the limit, where
$r \rightarrow 2M^+$. In this limit we have:
\begin{equation}\label{eq:2.21}
dm(r) = \frac{1}{8\pi}a\,dA.
\end{equation}
Indeed, just outside of the event horizon we have, in the leading
approximation:
\begin{subequations}
\begin{eqnarray}\label{eq:2.22a}
dA &=& d(16\pi M^2) = 32\pi M\,dM,\\
\label{eq:2.22b} a &=& \left(1 -
\frac{2M}{r}\right)^{-1/2}\frac{1}{4M}.
\end{eqnarray}
\end{subequations}
When these expressions are substituted in Eq. (2.21), we get:
\begin{equation}
dm(r) = \left(1 - \frac{2M}{r}\right)^{-1/2}\,dM,
\end{equation}
which is Eq. (2.15). For a distant observer $\frac{r}{M} \gg 1$ and
the increase in the mass of the hole is, from the point of view of
that observer, related to the increase $dA$ in the area $A$ of the
two-sphere, where $a = constant$, as:
\begin{equation}\label{eq:2.24}
dm(r) = \frac{1}{4\pi}a\,dA.
\end{equation}

Since the proper acceleration $a = constant$ during the formation of
the black hole, the expression
\begin{equation}\label{eq:2.25}
E_H = \frac{a}{8\pi}A
\end{equation}
may be regarded as the energy of the Schwarzschild black hole from
the point of view of an observer on the stretched horizon, whereas
from the point of view of a distant observer the energy is:
\begin{equation}\label{eq:2.26}
E_\infty = \frac{a}{4\pi}A,
\end{equation}
where $A = 4\pi r^2$ is the area of our two-sphere. Indeed, if one
puts $a = (1-\frac{2M}{r})^{-1/2}\frac{M}{r^2}$ in Eq. (2.26), one
finds that $E_\infty = M$, when $r\rightarrow\infty$. In other
words, our expression for energy coincides with the Schwarzschild
mass $M$, when the two-sphere $r = constant$ lies at the faraway
infinity.

\subsection{Microscopic Properties}

After having established in Eq. (2.25) an expression for the energy
$E_H$ of a Schwarzschild black hole from the point of view of an
observer on its stretched horizon in the spirit of the Brown-York
energy, we shall now turn our attention to the microscopic
properties of the stretched horizon. To put it simply, we shall
assume that the stretched horizon consists of a finite number of
discrete constituents, each of them contributing to the two-sphere
an area, which is an integer times a constant. As a consequence, the
area of the stretched horizon takes the form:
\begin{equation}\label{eq:2.27}
A = \alpha(n_1 + n_2 +...+ n_N),
\end{equation}
where $N$ is the number of the constituents, and $n_1, n_2,..., n_N$
are non-negative integers. $\alpha$ is a numerical constant to be
determined later. At this point we shall not specify what these
constituents actually are.  We simply say that the constituents have
independent {\it area eigenstates}, which are labelled by the
quantum numbers $n_j$ $(j = 1, 2,..., N)$, and the possible areas of
the stretched horizon are related to the quantum numbers $n_j$ as in
Eq. (2.27).  Of course, one could also attempt to write the stretched
horizon area as $A = \alpha_1 n_1 + \alpha_2 n_2 +...+ \alpha_N
n_N$, where the positive constants $\alpha_1$, $\alpha_2$,...,
$\alpha_N$ are all different, but that would be an unnecessary
complication, making the calculation of the partition function much
more~difficult.

One of the consequences of Eq. (2.27) is that the possible areas of
the stretched horizon are of the form
\begin{equation}\label{eq:2.28}
A_n = n\alpha,
\end{equation}
where
\begin{equation}\label{eq:2.29}
n := n_1 + n_2 +...+ n_N.
\end{equation}
In other words, the area of the stretched horizon has, in our model,
an equally spaced spectrum. Since our stretched horizon is assumed
to lie just outside of the event horizon of the Schwarzschild black
hole, the spectrum of Eq. (2.28) for the area of the stretched
horizon agrees with the area spectrum of the event horizon itself.
An equally spaced area spectrum for the event horizon area of a
black hole was first proposed by Jacob Bekenstein in 1974
\cite{seitseman}, and it was re-vitalized by Bekenstein and Mukhanov
in 1995 \cite{kahdeksan}. Since then, spectra of the form (2.28)
for the event horizon area have been obtained by several authors on
various grounds. (For a sample of papers on this subject see, for
example, Refs. [12-21]).

\section{The Partition Function}

Whenever one goes over from the microscopic to the macroscopic and
thermodynamical description of any system, the key role is played by
the {\it partition function}
\begin{equation}\label{eq:3.1}
Z(\beta) := \sum_n g(E_n)e^{-\beta E_n}
\end{equation}
of the system. In Eq. (3.1) $n$ labels the possible total energies
$E_n$ of the system, $\beta$ is its inverse temperature, and
$g(E_n)$ is the number of the degenerate states of the system
associated with the same total energy $E_n$. During the past 30
years or so it has become a commonly accepted wisdom that the
possible quantum states of a black hole are encoded (although nobody
knows exactly how) to the quantum states of its event horizon. In
other words---this hypothesis says---there should be a one-to-one
correspondence between the quantum states of a black hole, and those
of its event horizon [4-6]. Although it seems that nobody has
ever really proved the validity of this hypothesis, it is certainly
very useful, and in what follows, we shall always use it without
hesitation. Hence when talking about the quantum states of a black
hole, we actually talk about the quantum states of its event horizon
which, in turn, is approximated by the stretched horizon. As a
consequence, the partition function which we shall obtain in this
section for the Schwarzschild black hole is really the partition
function of its stretched horizon.

\subsection{Counting of States}

Using Eqs. (2.25) and (2.28), together with the assumptions
stated above, we find that the possible energies of the
Schwarzschild black hole from the point of view of an observer on
its stretched horizon are, according to our model, of the form:
\begin{equation}\label{eq:3.2}
E_n = n\alpha\frac{a}{8\pi},
\end{equation}
where $n = 0, 1, 2,...$, and $a$ is the proper acceleration of the
observer. In SI units, Eq. (3.2) takes the~form:
\begin{equation}\label{eq:3.3}
E_n = n\alpha\frac{\hbar a}{8\pi c},
\end{equation}
which implies that $\alpha$ is a pure, dimensionless constant.
Presumably, $\alpha$ is of the order of unity.

A more difficult problem is to find the the number $g(E_n)$ of the
degenerate states associated with the same total energy $E_n$ of the
hole. Basically, the problem is to identify the different
microscopic states of the stretched horizon of the hole. The
calculation of $g(E_n)$ is based on the following {\it statistical
postulate}:

\bigskip

{\it The microscopic states of the Schwarzschild black hole are
identified with the combinations of the non-vacuum area eigenstates
of the constituents of its stretched horizon.}

\bigskip

As the reader may recall, the area eigenstates of the constituents
of the stretched horizon are the states specified by the quantum
numbers $n_j$ ($j = 1, 2,..., N$), and we say that a constituent $j$
is in {\it vacuum}, if $n_j = 0$. Denoting the number of the
constituents by $N$, we find that according to our statistical
postulate the number $g(E_n)$ of the degenerate states of the hole
associated with the same total energy $E_n$ is the same as is the
number of different ways of expressing the positive integer $n$ as a
sum of at most $N$ positive integers. More precisely, it is the
number of ordered strings $(n_1,n_2,...,n_m)$, where $1\le m\le N$,
$n_j \in \lbrace 1, 2, 3,...\rbrace$ for all $j = 1, 2,..., m$, and
$n_1 + n_2 +...+ n_m = n$. It is important to note that different
orderings of the same quantum numbers represent, in our model,
different quantum states. Hence it follows that if we switch the
quantum states of two constituents while keeping the quantum states
of the other constituents as unchanged, the overall quantum state of
the constituents will also change. In this sense the constituents
have independent individual identities.

We shall now proceed to the calculation of $g(E_n)$. The number of
ways of writing a given positive integer $n$ as a sum of $m$
positive integers is the same as is the number of ways of arranging
$n$ balls is a row in $m$ groups by putting $(m - 1)$ identical
divisions in the $(n - 1)$ empty spaces between the balls. The
position for the first division may be chosen in $(n - 1)$ ways, for
the second in $(m - 2)$ ways, and so on. So the total number of the
combinations of the positions of the divisions is
\begin{equation}\label{eq:3.4}
(n - 1)(n - 2)\cdots(n-m+1) = \frac{(n-1)!}{(n-m)!}.
\end{equation}
However, since the divisions are identical, we must divide this
number by the number of the possible orderings of the divisions,
which is $(m - 1)(m - 2)\cdots 2\cdot 1 = (m - 1)!$. Hence the
number of ways of writing a positive integer $n$ as a sum of $m$
positive integers is given by the binomial coefficient
\begin{equation}\label{eq:3.5}
 \left(\begin{array}{cc}n-1\\m-1\end{array}\right) = \frac{(n-1)!}{(m-1)!(n-m)!}.
\end{equation}
For instance, the number of ways of writing the number $5$ as a sum
of $3$ positive integers is
\begin{equation}\label{eq:3.6}
 \left(\begin{array}{cc}4\\2\end{array}\right) = \frac{4!}{2!(4-2)!} = 6.
\end{equation}
Indeed, we have:
\begin{equation}\label{eq:3.7}
5 = 1 + 1 + 3 = 1 + 3 + 1 = 3 + 1 + 1 = 2 + 2 + 1 = 2 + 1 + 2 = 1 +
2 + 2.
\end{equation}
The considerations performed above imply that the degeneracy of a
state with energy $E_n$ is
\begin{equation}\label{eq:3.8}
g(E_n) = \sum_{m=1}^N
\left(\begin{array}{cc}n-1\\m-1\end{array}\right),
\end{equation}
whenever $N \le n$. In the special case, where $N = n$, we have
\begin{equation}\label{eq:3.9}
g(E_n) = \sum_{m=1}^n
\left(\begin{array}{cc}n-1\\m-1\end{array}\right) = 2^{n-1}.
\end{equation}
If $N > n$, $g(E_n)$ is simply the number of ways of expressing $n$
as a sum of positive integers, no matter how many. Since the maximum
number of those positive integers is $n$, we find that $g(E_n)$ is
given by Eq. (3.9), whenever $N \ge n$.

\subsection{The Partition Function}

After finding $g(E_n)$ we are now able to write the partition
function $Z(\beta)$ of the Schwarzschild black hole from the point
of view of an observer on its stretched horizon. Using Eqs. (3.1),
(3.3), (3.8) and (3.9) we get:
\begin{equation}\label{eq:3.10}
Z(\beta) = Z_1(\beta) + Z_2(\beta),
\end{equation}
where
\begin{subequations}
\begin{eqnarray}\label{eq:3.11a}
Z_1(\beta) &:=& \sum_{n=1}^N 2^{n-1}e^{-n\beta E_1},\\
\label{eq:3.11b} Z_2(\beta) &:=&
\sum_{n=N+1}^\infty\left\lbrack\sum_{k=0}^N
\left(\begin{array}{cc}n-1\\k\end{array}\right) e^{-n\beta
E_1}\right\rbrack.
\end{eqnarray}
\end{subequations}
It turns out useful to define the temperature
\begin{equation}\label{eq:3.12}
T_C := \frac{\alpha\hbar a}{8(\ln 2)\pi k_Bc}.
\end{equation}
When written in terms of $T_C$, $Z_1(\beta)$ and $Z_2(\beta)$ take,
in the natural units, the forms:
\begin{subequations}
\begin{eqnarray}\label{eq:3.13a}
Z_1(\beta) &=& \frac{1}{2}\sum_{n=1}^N 2^{(1-\beta T_C)n},\\
\label{eq:3.13b} Z_2(\beta) &=&
\sum_{n=N+1}^\infty\left\lbrack\sum_{k=0}^N
\left(\begin{array}{cc}n-1\\k\end{array}\right) 2^{-n\beta
T_c}\right\rbrack.
\end{eqnarray}
\end{subequations}
We shall see later that the temperature $T_C$ will play an important
role in the statistical and the thermodynamical properties of the
Schwarzschild black hole. We shall call $T_C$ as the {\it
characteristic temperature} of the Schwarzschild black hole.

The calculation of the partition function $Z(\beta)$ has been
performed in details in Appendix A. It is most gratifying that the
calculations may be performed explicitly. The result turns out to be
surprisingly simple. We find:
\begin{equation}\label{eq:3.14}
Z(\beta) = \frac{1}{2^{\beta T_C} - 2}\left\lbrack 1 -
\left(\frac{1}{2^{\beta T_C} - 1}\right)^{N+1}\right\rbrack,
\end{equation}
if $\beta T_C\ne 1$, and
\begin{equation}\label{eq:3.15}
Z(\beta) = N + 1,
\end{equation}
if $\beta T_C = 1$.

\section{Energy \emph{vs.} Temperature}
After finding in Eq. (3.14) an explicit expression for the partition
function of the Schwarzschild black hole from the point of view of
an observer on its stretched horizon, we are now prepared to obtain
expressions for various thermodynamical quantities of the hole. The
first of them is the average energy
\begin{equation}\label{eq:4.1}
E(\beta) = -\frac{\partial}{\partial\beta}\ln Z(\beta)
\end{equation}
of the hole in a given temperature $T = \frac{1}{\beta}$. Using
Eq. (3.14) we find:
\begin{equation}\label{eq:4.2}
E(\beta) = \left\lbrack\frac{2^{\beta T_C}}{2^{\beta T_C} - 2} -
\frac{(N+1)2^{\beta T_C}}{(2^{\beta T_C} - 1)^{N+2} - 2^{\beta T_C}
+ 1}\right\rbrack T_C\ln 2.
\end{equation}
It turns out useful to define the average energy of the hole per a
constituent:
\begin{equation}\label{eq:4.3}
\bar{E}(\beta) := \frac{E(\beta)}{N}
\end{equation}
and we get, assuming that $N$ is very large:
\begin{equation}\label{eq:4.4}
\bar{E}(\beta) = \bar{E}_1(\beta) + \bar{E}_2(\beta),
\end{equation}
where
\begin{subequations}
\begin{eqnarray}\label{eq:4.5a}
\bar{E}_1(\beta) &:=& \frac{1}{N}\frac{2^{\beta T_C}}{2^{\beta T_C} - 2}T_C\ln 2,\\
\label{eq:4.5b} \bar{E}_2(\beta) &:=& -\frac{2^{\beta
T_C}}{(2^{\beta T_C} - 1)^{N+2} - 2^{\beta T_C} + 1}T_C\ln 2.
\end{eqnarray}
\end{subequations}
When obtaining Eq. (4.4) we have approximated $(N+1)/N$ by 1. It
should be noted that the quantity $\bar{E}(\beta)$ may not be
interpreted as the average energy of an individual constituent of
the stretched horizon. The constituents of the stretched horizon are
presumably Planck size objects, and at the Planck length scales the
concept of energy simply does not make sense. However, using the
quantity $\bar{E}(\beta)$ we may consider the distribution of the
constituents on different quantum states as a function of the
inverse temperature $\beta$. Using Eqs. (3.3), (3.12) and
(4.3) we find that the average value
\begin{equation}\label{eq:4.6}
\bar{n} := \frac{n_1 + n_2 +...+ n_N}{N}
\end{equation}
of the quantum numbers $n_j$ determining the quantum states of
individual constituents is related to $\bar{E}(\beta)$ such that
\begin{equation}\label{eq:4.7}
\bar{n}(\beta) = \frac{\bar{E}(\beta)}{T_C\ln 2}.
\end{equation}

Since the constituents of the stretched horizon are presumably
Planck size objects, one may expect that for real, astrophysical
black holes the number $N$ of the constituents is enormous. For
instance, if the mass of a black hole is a few solar masses, its
Schwarzschild radius is a few kilometers, and $N$ is around
$10^{77}$. When investigating the properties of the stretched
horizon we may therefore consider, in practice, the limit where $N$
goes to infinity. In this limit the properties of the quantity
$\bar{E}(\beta)$ depend crucially on whether the absolute
temperature $T$ of the stretched horizon is greater or less than the
characteristic temperature $T_C$.

If $T < T_C$, the quantity $\beta T_C > 1$ in the natural units. As
a consequence, $\bar{E}_1(\beta)$ is positive, and it vanishes in
the limit, where $N\rightarrow\infty$. One also finds that the
quantity $2^{\beta T_C} - 1 > 1$, and hence the quantity $(2^{\beta
T_C} - 1)^{N+2}$ in the denominator of $\bar{E}_2(\beta)$ goes
towards the positive infinity, when $N\rightarrow\infty$. So we find
that $\bar{E}_2(\beta)$ will vanish in this limit as well, and we
get an important result:
\begin{equation}\label{eq:4.8}
\lim_{N\rightarrow\infty}\bar{E}(\beta) = 0,
\end{equation}
whenever $T < T_C$. This means that all constituents of the
stretched horizon are, in average, in vacuum, when $T < T_C$. When
$T = 0$, we must consider the limit, where both $\beta T_C$ and $N$
go towards the positive infinity. In this limit the first term
inside the brackets in Eq. (4.2) goes towards unity, whereas the
second term will vanish. Hence we find that when $T = 0$, the total
energy of the hole is, in SI units, $E = k_B T_C\ln 2$. The fact
that the energy is non-zero even, when $T = 0$ implies that the
Schwarzschild black hole may never, according to our model, vanish
completely, but at least a Planck size remnant is left in behind.
The result is a straightforward consequence of the statistical
postulate, which implies that at least one of the constituents of
the stretched horizon is in a non-vacuum state. It is possible to
construct well-defined quantum-mechanical models of the
Schwarzschild black hole, where the ground state energy is non-zero
\cite{yhdeksan4}. The possible evaporation of the Planck size
remnant, together with the consequent solution of the information
loss problem in black hole physics has been considered in
\cite{kymmenen1, kymmenen2}.

When $T > T_C$, the quantity $\beta T_C < 1$ and $0 < 2^{\beta T_C}
- 1 < 1$. As a consequence, the quantity $(2^{\beta T_C} - 1)^{N+2}$
in the denominator of $\bar{E}_2(\beta)$ will vanish in the limit,
where $N\rightarrow\infty$. The quantity $\bar{E}_1(\beta)$ will
also vanish in this limit, and hence it follows from Eqs. (4.4),
(4.5a) and (4.5b) that we may write, in~effect,
\begin{equation}\label{eq:4.9}
\bar{E}(\beta) = \frac{2^{\beta T_C}}{2^{\beta T_C} - 1}T_C\ln 2,
\end{equation}
whenever $T > T_C$.

Of particular interest is the high temperature limit, where the
absolute temperature $T \gg T_C$. In this limit $\beta T_C$ is very
small, and because
\begin{equation}\label{eq:4.10}
2^{\beta T_C} = 1 + \beta T_C\ln 2 + {\cal O}((\beta T_C)^2),
\end{equation}
where ${\cal O}((\beta T_C)^2)$ denotes the terms, which are of the
order $(\beta T_C)^2$, or higher, Eq. (4.9) implies:
\begin{equation}\label{eq:4.11}
\bar{E}(\beta) = \frac{1}{\beta} + {\cal O}(1)
\end{equation}
where ${\cal O}(1)$ denotes the terms, which are of the order
$(\beta T_C)^0$ or higher. Hence we find, using Eqs. (4.3) and
(4.9), that in the high temperature limit the energy $E$ of the
Schwarzschild black hole from the point of view of an observer on
its stretched horizon takes, in SI units, the form:
\begin{equation}\label{eq:4.12}
E(T) = Nk_B T.
\end{equation}
So we have managed to obtain a result, which holds for almost any
system in a high enough temperature. The thermal energy of almost
any system may be written, in a high temperature limit, in the form:
\begin{equation}\label{eq:4.13}
E(T) = \gamma Nk_BT,
\end{equation}
where $N$ is the number of the constituents of the system, and
$\gamma$ is a pure number, which depends on the number of the
physical degrees of freedom possessed by an individual constituent
of the system. For instance, the solids obey, as a very good
approximation, the Dulong-Petit law \cite{kymmenen}:
\begin{equation}\label{eq:4.14}
E(T) = 3Nk_BT,
\end{equation}
where $N$ is the number of the constituents (atoms, molecules or
ions) of the solid. In our model each constituent of the stretched
horizon possesses exactly one physical degree of freedom, which is
described by the quantum number $n_j$ $(j = 1, 2,..., N)$. Hence
Eq. (4.12) is something one might expect, and it may therefore be
used as a consistency check of our model.

  \section{Phase Transition and the Hawking Effect}

So far we have considered the properties of the average energy of
the Schwarzschild black hole from the point of view of an observer
on its stretched horizon, when the temperature $T$ of the hole is
either smaller or greater than the characteristic temperature $T_C$.
When the temperature of the hole is very close to the characteristic
temperature $T_C$, something very peculiar happens to its energy.

It has been shown in Appendix B that when $T = T_C$, the average
energy per a constituent of the stretched horizon is, in SI units,
\begin{equation}\label{eq:5.1}
\bar{E} = k_BT_C\ln 2,
\end{equation}
and that
\begin{equation}\label{eq:5.2}
\frac{d\bar{E}}{dT}\vert_{T=T_C} = \frac{1}{6}k_B(\ln 2)^2N +  {\cal
O}(1),
\end{equation}
where ${\cal O}(1)$ denotes the terms, which are of the order $N^0$,
or less. Hence we observe that when the number $N$ of the
constituents becomes very large, $\frac{d\bar{E}}{dT}\vert_{T=T_C}$
goes towards the positive infinity. Putting in another way, this
means that increase in energy does not change the temperature of the
Schwarzschild black hole, when $T = T_C$. In other words, the
Schwarzschild black hole undergoes a {\it phase transition} at the
characteristic temperature $T = T_C$. Putting $T = T_C$ in Eq. (4.9),
which gives the quantity $\bar{E}(\beta)$ in the large $N$ limit,
whenever $T > T_C$, we find that during this phase transition
$\bar{E}(\beta)$ jumps from zero to the value
\begin{equation}\label{eq:5.3}
\bar{L} := 2k_B T_C\ln 2,
\end{equation}
which gives the latent heat per a constituent in the phase
transition.

The results obtained above are confirmed by numerical
investigations. In Figure 1 we have made a plot of $\bar{E}$ as a
function of the absolute temperature $T$, when $N = 100$. When $T <
T_C$, $\bar{E}$ is practically zero. However, when $T = T_C$, the
curve $\bar{E} = \bar{E}(T)$ becomes practically vertical. When $T$
is slightly greater than $T_C$, $\bar{E}(T)$ is approximately $1.4
k_BT_C$, which is about the same as $2\ln 2$. Finally, the
dependence of $\bar{E}(T)$ on $T$ becomes approximately linear, when
$T \gg T_C$.
\begin{figure}[htb!]
\begin{center}
\includegraphics{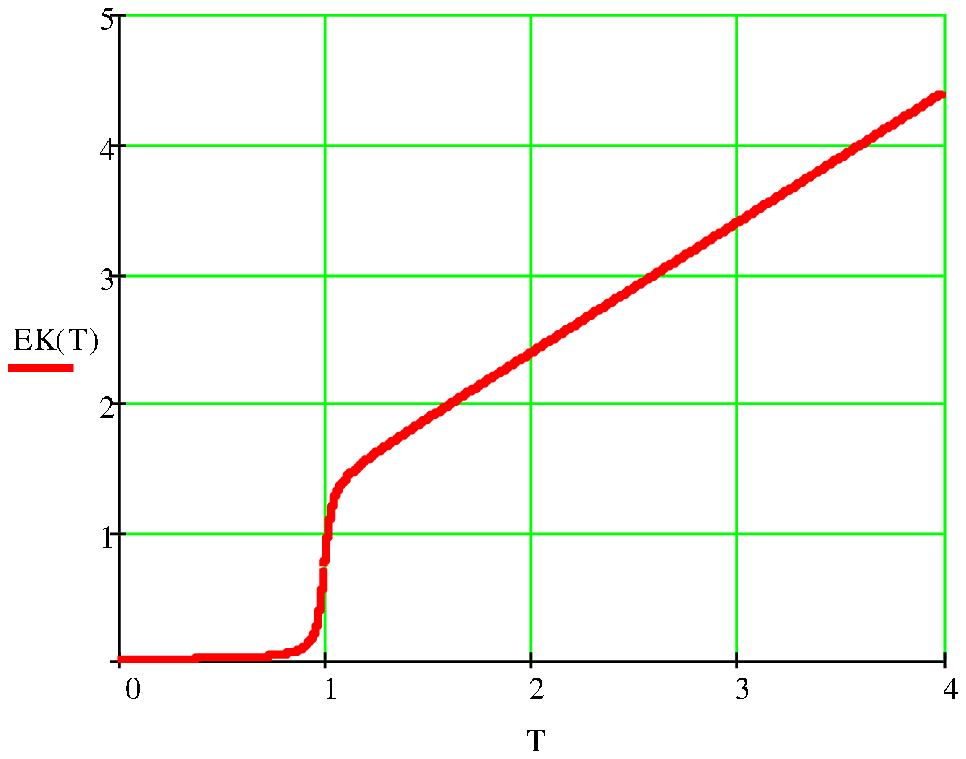}
\caption{The average energy ${\bar E}$ $(= EK(T))$ of the
Schwarzschild black hole per a constituent of its stretched horizon
as a function of the absolute temperature $T$, when the number of
the constituents of the stretched horizon is $N = 100$. The absolute
temperature $T$ has been expressed in the units of $T_C$, and the
average energy ${\bar E}$ in the units of $k_BT_C$. If $T < T_C$,
${\bar E}$ is effectively zero, which means that the constituents of
the stretched horizon (except one) are in vacuum. When $T = T_C$,
the curve ${\bar E} = {\bar E}(T)$ is practically vertical, which
indicates a phase transition at the temperature $T = T_C$. During
this phase transition the constituents of the stretched horizon are
excited from the vacuum to the second excited states. The latent
heat per a constituent corresponding to this phase transition is
${\bar L} = 2(\ln 2)k_BT_C \approx 1.4k_BT_C$. When $T > T_C$, the
curve ${\bar E} = {\bar E}(T)$ is approximately linear.
  \label{fig:Figure1}}
\end{center}
\end{figure}

The most important implication of the observed phase transition at
the characteristic temperature $T_C$ is that it {\it predicts the
Hawking effect}: The result that $\bar{E}(T)$ is practically zero,
when $T < T_C$, and then suddenly jumps to $\bar{L} = 2k_B T_C\ln
2$, when $T = T_C$, indicates that the characteristic temperature
$T_C$ is the lowest possible temperature a black hole may have. If
the temperature $T$ of the black hole were less than its
characteristic temperature $T_C$, all of the constituents of its
stretched horizon, except one, would be in vacuum, and there would
be no black hole. Using Eqs. (2.5) and (3.12) we find that the
characteristic temperature $T_C$ may be written in terms of the
Schwarzschild mass $M$ and the Schwarzschild radial coordinate $r$
of an observer on the stretched horizon as:
\begin{equation}\label{eq:5.4}
T_C = \frac{\alpha}{8\pi\ln 2}\left(1 -
\frac{2M}{r}\right)^{-1/2}\frac{M}{r^2}.
\end{equation}
On the stretched horizon $r$ is approximately $2M$, and hence an
observer just outside of the event horizon measures a temperature
\begin{equation}\label{eq:5.5}
T_C = \frac{\alpha}{32\pi\ln 2}\left(1 -
\frac{2M}{r}\right)^{-1/2}\frac{1}{M}
\end{equation}
for the black hole. As a consequence of the non-zero temperature of
the hole, thermal radiation comes out of the hole, and if the
possible backscattering effects of the radiation from the spacetime
geometry are neglected, the Tolman relation \cite{nelja} implies
that an observer at the asymptotic space-like infinity measures for
the radiation a temperature
\begin{equation}\label{eq:5.6}
T_\infty = \frac{\alpha}{32\pi\ln 2}\frac{1}{M}
\end{equation}
or, in SI units,
\begin{equation}\label{eq:5.7}
T_\infty = \frac{\alpha}{32\pi\ln 2}\frac{\hbar
c^3}{k_BG}\frac{1}{M}.
\end{equation}
Hence it follows that if we fix the so far undetermined constant
$\alpha$ to be
\begin{equation}\label{eq:5.8}
\alpha = 4\ln 2,
\end{equation}
then
\begin{equation}\label{eq:5.9}
T_\infty = T_H,
\end{equation}
where
\begin{equation}\label{eq:5.10}
T_H := \frac{1}{8\pi}\frac{\hbar c^3}{k_BG}\frac{1}{M}
\end{equation}
is the Hawking temperature of the hole \cite{yksi}. So we have
managed to show that according to our model the Schwarzschild black
hole has, from the point of view of a distant observer, a certain
non-zero temperature which, with the choice (5.8) for the
constant $\alpha$, agrees with the Hawking temperature $T_H$ of the
hole. In this sense our model predicts the Hawking effect. One may
consider the determination of the constant $\alpha$ from the
requirement that the model must reproduce the Hawking effect, rather
than from an {\it ab initio} calculation as a weakness of our model.
However, such an {\it ab initio} calculation is unattainable as far
as the proper quantum theory of gravity is still lacking.

If the temperature of the environment of a black hole is, from the
point of view of an observer on the stretched horizon, less than the
characteristic temperature $T_C$, the hole begins to emit thermal
radiation to its surroundings. As a consequence, the black hole
evaporates away, and our model enables us to investigate what
happens during this evaporation.

Suppose that at the onset of the evaporation the temperature of the
hole is slightly greater than its characteristic temperature $T_C$.
In that case the energy $\bar{E}$ per a constituent of the stretched
horizon agrees with the latent heat $\bar{L}$ per a constituent
given by Eq. (5.3). Since the average value $\bar{n}$ taken by the
quantum numbers $n_j$ is related to $\bar{E}$ as in Eq. (4.7), we
find that
\begin{equation}\label{eq:5.11}
\bar{n} = 2
\end{equation}
at the onset of the evaporation. This means that when the
evaporation begins, all of the constituents of the stretched horizon
are, in average, on the second excited state. During the evaporation
the constituents descent to lower quantum states until, finally, all
constituents, except one, are in vacuum, and the only remaining
constituent is in the first excited state. It should be noted that
from the point of view of an observer on the stretched horizon the
temperature of the hole remains the same during the whole process of
evaporation. That is because the proper acceleration $a$ of our
observer is, by definition, a constant, and the characteristic
temperature $T_C$ depends on the proper acceleration $a$ according
to Eq. (3.12). Hence we find that constant proper acceleration
implies constant temperature, and there is not any dramatic increase
in the black hole temperature during the final stages of its
evaporation. The temperature remains the same all the time, and the
black hole simply fades away, leaving a Planck-size remnant
in~behind.

\section{Entropy \emph{vs.} Horizon Area}

If we know the partition function $Z(\beta)$ of a system, we are
able to calculate its entropy. In general, the energy $E$, absolute
temperature $T$ and the entropy $S$ of any system obey the
relationship:
\begin{equation}\label{eq:6.1}
F = E - TS,
\end{equation}
where
\begin{equation}\label{eq:6.2}
F := -k_BT\ln Z
\end{equation}
is the Helmholtz free energy of the system. Hence the entropy of the
system may be written, in the natural units, as:
\begin{equation}\label{eq:6.3}
S(\beta) = \beta E(\beta) + \ln Z(\beta).
\end{equation}
Using Eqs. (3.14) and (4.2) we therefore find that according to
our model, the entropy of the Schwarzschild black hole when written
in terms  of its inverse temperature $\beta$ takes the form:
\begin{eqnarray}\label{eq:6.4}
S(\beta) &= \left\lbrack\frac{2^{\beta T_C}}{2^{\beta T_C} - 2} -
\frac{(N+1)2^{\beta T_C}}{(2^{\beta T_C} - 1)^{N+2} - 2^{\beta T_C} + 1}\right\rbrack\beta T_C\ln 2\nonumber \\
&+ \ln\left\lbrace\frac{1}{2^{\beta T_C} - 2}\left\lbrack 1 -
\left(\frac{1}{2^{\beta T_C} - 1}\right)^{N+1}
\right\rbrack\right\rbrace.
\end{eqnarray}

One immediately observes that in the limit, where $T\rightarrow 0$
and hence $\beta\rightarrow\infty$, the entropy of the hole
vanishes:
\begin{equation}\label{eq:6.5}
\lim_{T\rightarrow 0}S(T) = 0,
\end{equation}
which means that the black hole obeys the third law of
thermodynamics. More generally, it turns out useful to define a
quantity
\begin{equation}\label{eq:6.6}
\bar{S}(\beta) := \frac{\bar{S}(\beta)}{N},
\end{equation}
which gives the entropy per a constituent of the stretched horizon.
We find:
\begin{equation}\label{eq:6.7}
\bar{S}(\beta) = \bar{S}_1(\beta) + \bar{S}_2(\beta)
\end{equation}
where, in the large $N$ limit:
\begin{subequations}
\begin{eqnarray}\label{eq:6.8a}
\bar{S}_1(\beta) &:=& \left\lbrack\frac{1}{N}\frac{2^{\beta
T_C}}{2^{\beta T_C} - 2}
- \frac{2^{\beta T_C}}{(2^{\beta T_C} - 1)^{N+2} - 2^{\beta T_C} + 1}\right\rbrack\beta T_C\ln 2,\\
\label{eq:6.8b} \bar{S}_2(\beta) &:=&
\frac{1}{N}\ln\left\lbrace\frac{1}{2^{\beta T_C} - 2}\left\lbrack 1
- \left(\frac{1}{2^{\beta T_C} -
1}\right)^{N+1}\right\rbrack\right\rbrace.
\end{eqnarray}
\end{subequations}
If $T < T_C$, then $2^{\beta T_C} - 1 > 1$, and the quantity
$(2^{\beta T_C} - 1)^N$ goes towards the positive infinity in the
large $N$ limit. As a consequence, we get:
\begin{equation}\label{eq:6.9}
\lim_{N\rightarrow\infty}\bar{S}(\beta) = 0,
\end{equation}
whenever $T < T_C$. If $T > T_C$, then $2^{\beta T_C} - 1 < 1$, and
the quantity $(2^{\beta T_C} - 1)^N$ goes towards zero in the large
$N$ limit. Hence we have:
\begin{equation}\label{eq:6.10}
\lim_{N\rightarrow\infty}\bar{S}(\beta) = \frac{2^{\beta
T_C}}{2^{\beta T_C} - 1}\beta T_C\ln 2 - \ln(2^{\beta T_C} - 1),
\end{equation}
whenever $T > T_C$. So we find that in the large $N$ limit there is
a discrete jump in the values taken by $\bar{S}(\beta)$ at the phase
transition temperature $T = T_C$, and the magnitude of this jump is
given by the right hand side of Eq. (6.10).

A really interesting question is in which way does the entropy $S$
of the Schwarzschild black hole depend on its event horizon area.
Eq. (4.9) implies that when $T > T_C$, the quantity $\beta T_C$
depends on the average energy $\bar{E}$ per a constituent of its
stretched horizon such that
\begin{equation}\label{eq:6.11}
\beta T_C = \frac{1}{\ln 2}\ln\left(\frac{\bar{E}}{\bar{E} - T_C\ln
2}\right),
\end{equation}
and on the energy $E = N\bar{E}$ of the hole as:
\begin{equation}\label{eq:6.12}
\beta T_C = \frac{1}{\ln 2}\ln\left(\frac{E}{E - NT_C\ln 2}\right).
\end{equation}
Employing Eqs. (2.25), (3.12) and (5.8) we find that the
stretched horizon area $A$ of the hole may be expressed in terms of
its energy $E$, in the natural units, as:
\begin{equation}\label{eq:6.13}
A = \frac{4E}{T_C},
\end{equation}
and hence
\begin{equation}\label{eq:6.14}
\beta T_C = \frac{1}{\ln 2}\ln\left(\frac{2A}{2A - A_{crit}}\right),
\end{equation}
where we have defined the {\it critical area}
\begin{equation}\label{eq:6.15}
A_{crit} := 8N\ln 2,
\end{equation}
which gives the area taken by the stretched horizon, when $T$ is
slightly greater than $T_C$. Indeed, Eq. (6.14) implies:
\begin{equation}\label{eq:6.16}
\lim_{A\rightarrow A^+_{crit}}(\beta T_C) = 1.
\end{equation}
When $A = A_{crit}$, the constituents of the stretched horizon are,
in average, in the second excited states. In SI units $A_{crit}$ may
be written in terms of the Planck length $\ell_{Pl}$ as:
\begin{equation}\label{eq:6.17}
A_{crit} = 8N\ell_{Pl}^2\ln 2.
\end{equation}
Substituting the right hand side of Eq. (6.14) for $\beta T_C$ in
Eq. (6.10) we find that in the large $N$ limit we have:
\begin{equation}\label{eq:6.18}
\bar{S}(\beta) = \frac{2A}{A_{crit}}\ln\left(\frac{2A}{2A -
A_{crit}}\right) + \ln\left(\frac{2A - A_{crit}}{A_{crit}}\right),
\end{equation}
whenever $A \ge A_{crit}$. Hence the entropy $S = N\bar{S}$ takes,
by means of Eq. (6.15), the form:
\begin{equation}\label{eq:6.19}
S(A) = \frac{1}{4\ln 2}A\ln\left(\frac{2A}{2A - A_{crit}}\right) +
N\ln\left(\frac{2A - A_{crit}}{A_{crit}}\right)
\end{equation}
or, in SI units:
\begin{equation}\label{eq:6.20}
S(A) = \frac{1}{4\ln 2}\frac{k_Bc^3}{\hbar G}A\ln\left(\frac{2A}{2A
- A_{crit}}\right) + Nk_B\ln\left(\frac{2A -
A_{crit}}{A_{crit}}\right).
\end{equation}

Since the stretched horizon of the Schwarzschild black hole lies
just outside of its event horizon, we may equate the stretched
horizon area $A$ of the hole with its event horizon area. It is
interesting to see what happens to the black hole entropy when $A =
A_{crit}$. When $A = A_{crit}$, the hole has just undergone the
phase transition, where the constituents of its stretched horizon
have, in average, jumped from the vacuum to the second excited
states, and the temperature $T$ of the hole is slightly above of its
characteristic temperature $T_C$. Putting $A = A_{crit}$ in
Eq. (6.20) we get:
\begin{equation}\label{eq:6.21}
S(A) = \frac{1}{4}\frac{k_Bc^3}{\hbar G}A
\end{equation}
which is exactly the Bekenstein-Hawking entropy law \cite{viisi}.
We have thus achieved one of our main goals: We have been able to
obtain the Bekenstein-Hawking entropy law for the Schwarzschild
black hole from its partition function which, in turn, followed from
a specific microscopic model of its stretched horizon. It should be
stressed, however, that our derivation of the Bekenstein-Hawking
entropy law holds only if the horizon area of the Schwarzschild
black hole agrees with its critical area $A_{crit}$ and the
temperature $T$ of the hole is slightly higher than its
characteristic temperature $T_C$. If $T$ is appreciably greater than
$T_C$, and thus $A$ is appreciably greater than $A_{crit}$, the
simple proportionality between the area and the entropy predicted by
the Bekenstein-Hawking entropy law for the Schwarzschild black hole
will no longer hold, and the Bekenstein-Hawking entropy law must be
replaced by Eq. (6.20). Since the characteristic temperature $T_C$
measured by an observer on the stretched horizon of the
Schwarzschild black hole corresponds to the Hawking temperature
$T_H$ of Eq. (5.10) measured by a faraway observer, we may thus
conclude that the Bekenstein-Hawking entropy law holds only if the
temperature of the hole is, from the point of view of a faraway
observer, very close to its Hawking temperature, but not otherwise.

So far we have managed to obtain, in Eq. (6.20), an expression for
the black hole entropy, when $T > T_C$ and $A \ge A_{crit}$. When $T
< T_C$, the constituents of its stretched horizon are effectively in
vacuum, and the black hole, as well as its entropy, will effectively
vanish. It is very interesting to investigate what will happen to
the black hole entropy during the phase transition where $T = T_C$
and its horizon area is {\it less} than its critical area
$A_{crit}$. It is a general property of any system that its entropy
$S$ is related to its energy $E$ and inverse temperature $\beta$
such that
\begin{equation}\label{eq:6.22}
\frac{\partial S}{\partial E} = \beta.
\end{equation}
Actually, this is the {\it definition} of the concept of temperature
in terms of the concepts of energy and entropy. Since the proper
acceleration $a$ of the observer is assumed to be a constant, we
get, using Eq. (2.25):
\begin{equation}\label{eq:6.23}
\frac{\partial S}{\partial E} = \frac{\partial S}{\partial
A}\frac{dA}{dE} = \frac{\partial S}{\partial A}\frac{8\pi}{a}.
\end{equation}
Eqs. (3.12) and (5.8) imply that when $T = T_C$, the inverse
temperature of the hole is, in natural units,
\begin{equation}\label{eq:6.24}
\beta = \frac{2\pi}{a},
\end{equation}
and hence it follows from Eqs. (6.22) and (6.23) that
\begin{equation}\label{eq:6.25}
\frac{\partial S}{\partial A} = \frac{1}{4},
\end{equation}
or
\begin{equation}\label{eq:6.26}
S = \frac{1}{4}A
\end{equation}
which, again, is the Bekenstein-Hawking entropy law in the natural
units. So we have managed to show that whenever $A \le A_{crit}$,
the entropy of the Schwarzschild black hole obeys the
Bekenstein-Hawking entropy law.

Before closing our discussion on the entropy of the Schwarzschild
black hole we point out yet another interesting feature of our
model. It has been shown in Appendix B that when $T = T_C$, the
energy of the hole from the point of view of an observer on its
stretched horizon is exactly \cite{yksitoista}
\begin{equation}\label{eq:6.27}
 E = (N + 2)k_BT_C\ln 2.
\end{equation}
Hence it follows from Eqs. (6.13) and (6.26) that when $T = T_C$,
the entropy of the Schwarzschild black hole may be written in terms
of $N$, the number of the constituents of the stretched horizon, as:
\begin{equation}\label{eq:6.28}
S = k_B\ln(2^{N+2}).
\end{equation}
Putting in another way, this means that when the temperature $T$ of
the hole is exactly the same as its characteristic temperature
$T_C$---which means that its temperature from the point of view of a
faraway observer agrees with its Hawking temperature $T_H$---each
constituent of the stretched horizon carries, in average, exactly
one bit of information. In this sense our model seems to reproduce
at least in some respects  Wheeler's famous ``it from bit''
proposal, which states in very broad terms that in the utmost
fundamental level, the laws of physics should be reducible to the
properties of some fundamental constituents each carrying exactly
one bit of information \cite{kaksitoista}.

\section{Discussion}

In this paper we have considered the statistical and the
thermodynamical properties of the Schwarzschild black hole from the
point of view of an observer on its ``stretched horizon'', or a
space-like two-sphere, where the Schwarzschild coordinate $r$ was
assumed to be slightly greater than the Schwarzschild radius $R_S =
2M$ of the hole. The stretched horizon was assumed to consist of a
finite number of discrete, Planck-size constituents, each of them
contributing an area, which is an integer times a constant, to the
total area of the stretched horizon. Assuming that the quantum
states of the Schwarzschild black hole are encoded in the quantum
states of the constituents of its stretched horizon, we wrote the
partition function of the hole. It turned out that the partition
function may be calculated explicitly, yielding a surprisingly
simple, analytic expression. Our partition function implied, among
other things, the Hawking effect, and the Bekenstein-Hawking
entropy law, which states that the black hole entropy is, in the
natural units, one-quarter of its event horizon area. The entropy of
the hole was found to agree with the Bekenstein-Hawking entropy,
when the temperature of the hole agrees, from the point of view of a
faraway observer, with the Hawking temperature of the hole. Using
our partition function, however, it is possible to obtain
expressions for the mass and the temperature of the hole even when
its temperature differs from its Hawking temperature. The Hawking
temperature is the lowest possible temperature of a black hole, but
if the hole is in a heat bath with a temperature higher than its
Hawking temperature, its entropy will differ from its
Bekenstein-Hawking entropy.

The most interesting feature of our model is that it provides a
microscopic interpretation to the Hawking effect: The Hawking effect
is a consequence of a {\it phase transition} performed by a black
hole. At a certain characteristic temperature $T_C$, which is
proportional to the proper acceleration $a$ of an observer on the
stretched horizon, the black hole undergoes a phase transition,
where the microscopic constituents of its stretched horizon descend,
in average, from the second excited states to the vacuum. During
this phase transition the black hole emits radiation with the
characteristic temperature $T_C$ until, finally, all constituents of
its stretched horizon, except one, are in vacuum, and there is no
more a black hole. Since the constituents of the stretched horizon
are in vacuum, whenever $T < T_C$, the characteristic temperature
$T_C$ is, from the point of view of an observer on the stretched
horizon, the lowest possible temperature a black hole may have. The
characteristic temperature involves a certain constant of
proportionality, and with an appropriate choice of that constant one
finds that at the asymptotic space-like infinity the temperature
$T_C$ corresponds to the Hawking temperature of the black hole. In
other words, the phase transition of the black hole takes the place,
from the point  of view of a faraway observer at rest with respect
to the hole, at its Hawking temperature, which is the same as is the
temperature of the radiation emitted by the hole during the phase
transition.

Even though our simple model of the Schwarzschild black hole meets
with some success, it deserves some critique as well. The crucial
point of our investigations was the counting of states. The counting
of states was based on what we called as ``statistical postulate''.
The statistical postulate implied that if we denote the number of
the constituents of the stretched horizon by $N$, then the number of
the microscopic states associated with the same total energy $E_n$
of a black hole is the same as is the number of the ordered strings
$(n_1,n_2,...,n_m)$  of positive integers $n_j$ $(j = 1, 2,.., m)$
such that $1 \le m \le N$ and $n_1 + n_2 +...+ n_m = n$. The
positive integers $n_j$ determine the quantum states of the
constituents of the stretched horizon, and it is important to note
that with this identification of microscopic states different
combinations of the {\it same} quantum numbers represent different
microscopic states. In this sense the counting of states in our
model is in marked contrast with the counting of states in, say, the
approaches to black hole thermodynamics based on loop quantum
gravity \cite{kolmetoista}, where different combinations of the same
quantum states of the individual constituents represent the same
microscopic state.

Indeed, the idea that different combinations of the same quantum
states of the individual constituents should represent the same
overall, microscopic state of a system might appear very attractive,
at least in the first sight. In that case the constituents of the
stretched horizon would behave in the same way as a system of
identical bosons, where interchange of two bosons keeps the overall
quantum state of the system unchanged. But then again, why should
the Planck scale constituents of the stretched horizon act like
identical bosons? The symmetry of the quantum state of a system of
identical bosons under interchanges of the bosons follows from the
Spin-Statistics Theorem which, in turn, may be traced back to the
symmetry properties of flat spacetime. There are no grounds to
believe that spacetime at the Planck length scale would possess
symmetries in any way akin to those of flat spacetime. Rather, the
contrary is the case. When constituents in different quantum states
are interchanged, the properties of the stretched horizon are also
changed. Hence there are no grounds to believe that the constituents
of the stretched horizon would behave like bosons either.

Another piece of criticism against our model may be addressed on the
form of our partition function. In a well known work
\cite{neljatoista} based on the Euclidean path integral approach to
quantum gravity it was shown by Hawking that in the leading
approximation the partition function of the Schwarzschild black hole
should be, in the natural units, of the form:
\begin{equation}\label{eq:7.1}
Z(\beta) = {\cal N}\exp(-\frac{\beta^2}{16\pi}),
\end{equation}
where ${\cal N}$ is an appropriate normalization constant. The
problem is that our partition function in Eq. (3.14) seems to be
nowhere near of the partition function (7.1) obtained by
Hawking. Is our partition function totally wrong?

To begin with, we note that in Eq. (3.14) the partition function has
been written from the point of view of an observer on the stretched
horizon, whereas the partition function in Eq. (7.1) has been written
from the point of view of an observer at rest at a faraway infinity.
Hence there is really nothing strange in the fact that the partition
functions (3.14) and (7.1) are very different. It is simply
what one~expects.

Nevertheless, there does exist a connection between the partition
functions (3.14) and (7.1). Since the energy of any system is
$E = -\frac{\partial}{\partial\beta}\ln Z(\beta)$, we find that
Eq. (7.1) may be written equivalently as:
\begin{equation}\label{eq:7.2}
\beta = 8\pi E.
\end{equation}
From the point of view of a faraway observer at rest with respect to
the black hole the energy $E$ may be identified with the
Schwarzschild mass $M$ of the hole, and hence Eq. (7.1) simply states
that Schwarzschild black hole has, in the leading approximation, the
temperature
\begin{equation}\label{eq:7.3}
T = \frac{1}{8\pi M},
\end{equation}
which is the Hawking temperature $T_H$ of the hole. However, this is
exactly what the partition function (3.14) is telling us as
well: It implies that even if we dropped the temperature of the
external heat bath of the hole close to the absolute zero, the hole
would still have a temperature which, from the point of view of a
faraway observer, equals with its Hawking temperature. We may
therefore view the right hand side of Eq. (7.1) as an effective
partition function of a black hole in the limit, where the
temperature of the external heat path is close to zero. An advantage
of the partition function (3.14) is that it enables us to
consider the thermodynamical properties of the Schwarzschild black
hole even when the temperature of the external heat bath exceeds the
Hawking temperature of the hole.

As we have learned, the characteristic temperature $T_C$ defined in
Eq. (3.12) plays a crucial role in our model. If we put $\alpha =
4\ln 2$ in Eq. (3.12), we find, in the natural units,
\begin{equation}\label{eq:7.4}
T_C = \frac{a}{2\pi}
\end{equation}
or, in SI units:
\begin{equation}\label{eq:7.5}
T_C = \frac{\hbar a}{2\pi k_Bc}.
\end{equation}
It is interesting that $T_C$ is exactly the {\it Unruh temperature}
measured by an observer with proper acceleration $a$
\cite{viisitoista}. The result suggests that a microscopic model
essentially similar to the one used for a microscopic interpretation
of the Hawking effect in this paper could possibly be used to
interpret the so-called Unruh effect as well. According to this
effect an observer in a uniformly accelerating motion will detect
particles even when all inertial observers detect a vacuum. The
effective temperature of the radiation of the particles measured by
an accelerating observer is the Unruh temperature, and hence the
Unruh temperature is the lowest possible temperature an accelerating
observer may detect in the same way as the Hawking temperature is
the lowest possible temperature of a black hole \cite{kuusitoista}.
One expects the emission of the Unruh radiation to involve processes
in the microscopic structure of spacetime, which are somewhat
similar to those taking place in the Hawking radiation.

To provide a microscopic interpretation to the Unruh effect one only
needs to replace the stretched horizon of the Schwarzschild black
hole by a spacelike two-surface propagating in spacetime close to
the Rindler horizon of an accelerated observer. In other words, the
stretched Schwarzschild horizon is replaced by a sort of ``stretched
Rindler horizon''. Postulating that any finite part with area $A$ of
the ``stretched Rindler horizon'' consists, like the stretched
Schwarzschild horizon, of a finite number of discrete constituents
and possesses, in the natural units, energy $E = \frac{a}{8\pi}A$
($a$ is the proper acceleration of the observer on the stretched
Rindler horizon), one finds that results identical to those obtained
for the stretched Schwarzschild horizon may be obtained for the
stretched Rindler horizon as well: At the temperature $T_U :=
\frac{a}{2\pi}$ the stretched Rindler horizon performs a phase
transition, where its constituents jump, in average, from the vacuum
to the second excited states, and the entropy of any finite part of
the stretched Rindler horizon is $S = \frac{1}{4}A$. Hence the Unruh
temperature $T_U = \frac{a}{2\pi}$ is the lowest temperature an
accelerated observer may measure, and we have found, in the context
of this simple model, a microscopic interpretation of the Unruh
effect. The most questionable step in our chain of reasoning was the
association of the concept of energy with the stretched Rindler
horizon in the same way as we associated the concept of energy with
the Schwarzschild black hole from the point of view of an observer
on the stretched Schwarzschild horizon. Lots of work must still be
done to clarify this point.

It seems that somewhat similar reasoning could be applied for an
extension of our analysis for Kerr-Newman black holes as well. As
the first step one finds such space-like two-surfaces just outside
of the event horizon of the Kerr-Newman black hole, where the
proper acceleration $a = constant$. A two-surface of this kind will
then serve as the stretched horizon of the hole. As the second step
one finds the constraint between the infinitesimal changes in the
mass $M$, electric charge $Q$ and the angular momentum $J$  of the
Kerr-Newman black hole such that no matter what may happen to the
hole, the proper acceleration $a$ on the stretched horizon will
always stay the same. The third step is to establish an expression,
beginning from the concept of Brown-York energy, for the energy of
the Kerr-Newman hole from the point of view of an observer on its
stretched horizon. If one is able to show that this energy takes, in
the natural units, the form $E = \frac{a}{8\pi}A$, where $A$ is the
area of the stretched horizon, the calculation of the partition
function of the Kerr-Newman black hole proceeds in the same way as
for the Schwarzschild black hole. One assumes that the stretched
horizon consists of $N$ discrete constituents such that its area may
be written in terms of non-negative integers $n_1, n_2,..., n_N$ as
in Eq. (1.4). The resulting partition function should be, when
written in terms of the inverse temperature $\beta$ and the constant
$\alpha$, the same as for the Schwarzschild black hole. One expects
this partition function to imply that from the point of view of an
observer at the space-like infinity the lowest possible temperature
of the hole is, in the natural units, $T = \frac{\kappa}{2\pi}$,
where $\kappa$ is the surface gravity on the event horizon, and that
the entropy of the hole is one-quarter of its event horizon area.
The technical details of the procedure outlined above, however, are
pretty complicated, and they will be left in the forthcoming
publications.

An essential feature of our model is its extreme simplicity. Indeed,
in our model the counting of the microscopic states, which is the
key question in the consideration of the statistical physics of any
system, boils down to the elementary problem of in how many ways a
given positive integer may be expressed as a sum of a given number
of integers. It remains to be seen whether any results of our model
will survive in the more advanced attempts to approach the problems
in the thermodynamics of black holes by means of the microphysics of
spacetime. An advantage of our model is that it allows us to address
these problems in precise terms.

\appendix
\section*{Appendix}
\section{Calculation of the Partition Function}

In this Appendix we derive the expression (3.14) for the
partition function $Z(\beta)$ of the Schwarzschild black hole.
Defining a quantity
\begin{equation}\label{eq:A.1}
q := 2^{-\beta T_C}
\end{equation}
one may write the sums $Z_1(\beta)$ and $Z_2(\beta)$ of Eqs. (3.13a)
and (3.13b) as:
\begin{subequations}
\begin{eqnarray}\label{eq:A.2a}
Z_1(\beta) &=& \frac{1}{2}\sum_{n=1}^N (2q)^n,\\
\label{eq:A.2b} Z_2(\beta) &=&
\sum_{n=N+1}^\infty\biggl[\sum_{k=0}^N
\left(\begin{array}{ccc}n-1\\k\end{array}\right)q^n\biggr].
\end{eqnarray}
\end{subequations}
Since $\beta$ and $T_C$ are both positive, $q < 1$. $Z_1(\beta)$ is
just a geometrical series, and we get:
\begin{equation}\label{eq:A.3}
Z_1(\beta) = q\frac{1 - (2q)^N}{1 - 2q},
\end{equation}
provided that $q \ne \frac{1}{2}$. If $q = \frac{1}{2}$, we have
\begin{equation}\label{eq:A.4}
Z_1(\beta) = \frac{1}{2}N.
\end{equation}

$Z_2(\beta)$ is much more difficult to calculate than $Z_1(\beta)$.
When calculating $Z_2(\beta)$, one of the key ideas is to write the
right hand side of Eq. (A.2b) by means of the higher order
derivatives of an appropriate function of $q$. Because, in general,
an arbitrary binomial coefficient may be written as:
\begin{equation}\label{eq:A.5}
\left(\begin{array}{ccc}n\\k\end{array}\right) =
\frac{1}{k!}n(n-1)(n-2)...(n-k+1),
\end{equation}
whenever $k > 0$, and
$\left(\begin{array}{ccc}n\\k\end{array}\right) = 1$, when $k = 0$,
one obtains a general formula
\begin{equation}\label{eq:A.6}
\left(\begin{array}{ccc}n\\k\end{array}\right)q^m =
\frac{1}{k!}q^{m-n+k}\frac{d^k}{dq^k}q^n,
\end{equation}
which yields:
\begin{equation}\label{eq:A.7}
\left(\begin{array}{ccc}n-1\\k\end{array}\right)q^n =
\frac{1}{k!}q^{k+1}\frac{d^k}{dq^k}q^{n-1}.
\end{equation}
So we find:
\begin{equation}\label{eq:A.8}
Z_2(\beta) =
\sum_{n=N+1}^\infty\biggl[\sum_{k=0}^N\frac{1}{k!}q^{k+1}\frac{d^k}{dq^k}q^{n-1}\biggr].
\end{equation}
Since one of the sums has a finite number of terms, we may change
the order of the summation, and we~get:
\begin{equation}\label{eq:A.9}
Z_2(\beta) =
\sum_{k=0}^N\biggl[\frac{1}{k!}q^{k+1}\frac{d^k}{dq^k}(q^N\sum_{n=0}^\infty
q^n)\biggr].
\end{equation}
Because $\vert q\vert < 1$, the geometric sum on the right hand side
of Eq. (A.9) will converge, and we have:
\begin{equation}\label{eq:A.10}
Z_2(\beta) =
\sum_{k=0}^N\biggl[\frac{1}{k!}q^{k+1}\frac{d^k}{dq^k}(\frac{q^N}{1-q})\biggr].
\end{equation}
As one may observe, we have managed to reduce a double sum with an
infinite number of terms into a simple sum with a finite number of
terms.

As the next step we employ the following formula, which is a
consequence of the product rule of differentiation:
\begin{equation}\label{eq:A.11}
\frac{d^k}{dq^k}[f_1(q)f_2(q)] =
\sum_{m=0}^k\left(\begin{array}{ccc}k\\m\end{array}\right)f_1^{(k-m)}(q)
f_2^{(m)}(q)
\end{equation}
for arbitrary smooth functions $f_1(q)$ and $f_2(q)$. If we define:
\begin{subequations}
\begin{eqnarray}\label{eq:A.12a}
f_1(q) &:=& q^N,\\
\label{eq:A.12b} f_2(q) &:=& \frac{1}{1-q},
\end{eqnarray}
\end{subequations}
we have:
\begin{subequations}
\begin{eqnarray}\label{eq:A.13a}
f_1^{(k-m)}(q) &=& \frac{N!}{(N-k+m)!}q^{N-k+m},\\
\label{eq:A.13b} f_2^{(m)}(q) &=& m!(1-q)^{-m-1},
\end{eqnarray}
\end{subequations}
and therefore Eq. (A.10) takes the form:
\begin{equation}\label{eq:A.14}
Z_2(\beta) =
\frac{q^{N+1}}{1-q}\sum_{k=0}^N\biggl[\sum_{m=0}^k\frac{N!}{(k-m)!(N-k+m)!}(\frac{q}{1-q})^m\biggr].
\end{equation}
When obtaining Eq. (A.14) we have used the formula:
\begin{equation}\label{eq:A.15}
\left(\begin{array}{ccc}k\\m\end{array}\right) = \frac{k!}{m!(k-m)!}.
\end{equation}
Using Eq. (A.15) we find:
\begin{equation}\label{eq:A.16}
Z_2(\beta) =
\frac{x^{N+1}}{(1+x)^N}\sum_{k=0}^N\biggl[\sum_{m=0}^k\left(\begin{array}{ccc}N\\k-m\end{array}\right)
x^m\biggr],
\end{equation}
where we have defined a new variable
\begin{equation}\label{eq:A.17}
x := \frac{q}{1-q}.
\end{equation}
Because $0 < q < 1$, $x$ is positive.

Now, it is possible to re-arrange the sums on the right hand side of
Eq. (A.16). As a result we get:
\begin{equation}\label{eq:A.18}
Z_2(\beta) =
\frac{x^{N+1}}{(1+x)^N}\sum_{n=0}^N\biggl[\left(\begin{array}{ccc}N\\n\end{array}\right)
\sum_{k=0}^{N-n}x^k\biggr].
\end{equation}
Because
\begin{equation}\label{eq:A.19}
\sum_{k=0}^{N-n}x^k = \frac{1 - x^{N-n+1}}{1-x},
\end{equation}
when $x \ne 1$, and
\begin{equation}\label{eq:A.20}
\sum_{k=0}^{N-n} x^k = N - n + 1,
\end{equation}
when $x = 1$, we have:
\begin{equation}\label{eq:A.21}
Z_2(\beta) =
\frac{x^{N+1}}{(1+x)^N}\frac{1}{1-x}\biggl[\sum_{n=0}^N\left(\begin{array}{ccc}N\\n\end{array}\right)
-
x^{N+1}\sum_{n=0}^N\left(\begin{array}{ccc}N\\n\end{array}\right)x^{-n}\biggr],
\end{equation}
when $x \ne 1$, and
\begin{equation}\label{eq:A.22}
Z_2(\beta) =
\frac{1}{2^N}\sum_{m=0}^N\biggl[\left(\begin{array}{ccc}N\\n\end{array}\right)(N
- n + 1)\biggr],
\end{equation}
when $x = 1$. Using the formulas: \cite{seitsemantoista}
\begin{subequations}
\begin{eqnarray}\label{eq:A.23a}
\sum_{n=0}^N\left(\begin{array}{ccc}N\\n\end{array}\right) &=& 2^N,\\
\label{eq:A.23b}
\sum_{n=0}^N n\left(\begin{array}{ccc}N\\n\end{array}\right) &=& N2^{N-1},\\
\label{eq:A.23c}
\sum_{n=0}^N\left(\begin{array}{ccc}N\\n\end{array}\right)(\frac{1}{x})^n
&=& \biggl(1 + \frac{1}{x}\biggr)^N,
\end{eqnarray}
\end{subequations}
we get:
\begin{equation}\label{eq:A.24}
Z_2(\beta) = \frac{x}{1-x}\biggl[(\frac{2x}{1+x})^N - x^{N+1}\biggr],
\end{equation}
when $x \ne 1$, and
\begin{equation}\label{eq:A.25}
Z_2(\beta) = \frac{1}{2}N + 1,
\end{equation}
when $x = 1$. When written in terms of the variable $q$, Eq. (A.24)
takes the form:
\begin{equation}\label{eq:A.26}
Z_2(\beta) = \frac{q^{N+1}}{1 - 2q}\biggl[2^N -
\frac{q}{(1-q)^{N+1}}\biggr].
\end{equation}
Combining Eqs. (A.2a), (A.2b), (A.3),(A.4), and (A.26) we get, when $\beta
\ne \frac{1}{T_C}$:
\begin{equation}\label{eq:A.27}
Z(\beta) = \frac{q}{1 - 2q}\biggl[1 -
\biggl(\frac{q}{1-q}\biggr)^{N+1}\biggr],
\end{equation}
and
\begin{equation}\label{eq:A.28}
Z(\beta) = N + 1,
\end{equation}
when $\beta = \frac{1}{T_C}$. Using Eq. (A.1) we find the final
expression for the partition function, when $\beta \ne
\frac{1}{T_C}$:
\begin{equation}\label{eq:A.29}
Z(\beta) = \frac{1}{2^{\beta T_C} - 2}\biggl[1 -
\biggl(\frac{1}{2^{\beta T_C} - 1}\biggr)^{N+1}\biggr],
\end{equation}
which is Eq. (3.14).

\section{Properties of the Partition Function Near the Characteristic Temperature}

In this Appendix we consider the energy of a black hole when the
absolute temperature $T$ measured by an observer on its stretched
horizon is very close to the characteristic temperature $T_C$.

Our starting point is Eq. (3.14), which gives the precise expression
for the partition function $Z(\beta)$ of a black hole. Because
$2^{\beta T_C} = 2$, when $T = T_C$, we denote:
\begin{equation}\label{eq:B.1}
y := 2^{\beta T_C} - 2,
\end{equation}
and  Eq. (3.14) takes the form:
\begin{equation}\label{eq:B.2}
Z(y) = \frac{1}{y}[1 - (1+y)^{-N-1}].
\end{equation}
When $T$ is close to $T_C$, $y$ is close to zero. When $y$ is close
to zero we may write, using Newton's binomial theorem:
\begin{eqnarray}\label{eq:B.3}
(1 + y)^{-N - 1} &= 1 - (N+1)y + \frac{1}{2!}(N+1)(N+2)y^2 \nonumber \\
&- \frac{1}{3!}(N+1)(N+2)(N+3)y^3 + ...,
\end{eqnarray}
and we get the Taylor expansion of $Z(y)$ around the point, where $y
= 0$:
\begin{equation}\label{eq:B.4}
Z(y) = (N+1) - \frac{1}{2!}(N+1)(N+2)y +
\frac{1}{3!}(N+1)(N+2)(N+3)y^2-....
\end{equation}
Applying the chain rule and the result
\begin{equation}\label{eq:B.5}
\frac{dy}{d\beta} = T_C(\ln 2)2^{\beta T_C}
\end{equation}
we find:
\begin{equation}\label{eq:B.6}
E(\beta) = -\frac{\partial}{\partial\beta}\ln Z(\beta) =
-\frac{Z'(y)}{Z(y)}T_C(\ln 2)(y+2),
\end{equation}
where $Z(y)$ is given by Eq. (B.4) and
\begin{eqnarray}\label{eq:B.7}
Z'(y) &= -\frac{1}{2}(N+1)(N+2) + \frac{1}{3}(N+1)(N+2)(N+3)y \nonumber \\
&- \frac{1}{8}(N+1)(N+2)(N+3)(N+4)y^2 +....
\end{eqnarray}
One readily finds that
\begin{subequations}
\begin{eqnarray}\label{eq:B.8a}
Z(0) &=& N+1,\\
\label{eq:B.8b} Z'(0) &=& - \frac{(N+1)(N+2)}{2}
\end{eqnarray}
\end{subequations}
which implies:
\begin{equation}\label{eq:B.9}
E(\frac{1}{T_C}) = (N+2)T_C\ln 2.
\end{equation}
Therefore, for very large $N$:
\begin{equation}\label{eq:B.10}
{\bar E}(\frac{1}{T_C}) = T_C\ln 2,
\end{equation}
which is Eq. (5.1).

It is interesting to consider the derivative of ${\bar E}$ with
respect to $T$, when $T = T_C$. Using Eq. (B.6) we get:
\begin{equation}\label{eq:B.11}
\frac{dE}{dy} = \bigg\lbrace\biggl[-\frac{Z''(y)}{Z(y)} +
\biggl(\frac{Z'(y)}{Z(y)}\biggr)^2 \biggr](y+2) -
\frac{Z'(y)}{Z(y)}\bigg\rbrace T_C\ln 2,
\end{equation}
and Eq. (B.4) implies:
\begin{equation}\label{eq:B.12}
\frac{dE}{dT}\vert_{T=T_C} = \frac{(\ln 2)^2}{6}(N + 2)(N + 3),
\end{equation}
where we have used the result:
\begin{equation}\label{eq:B.13}
\frac{dy}{dT} = -\frac{(\ln 2)T_C}{T^2}2^{\beta T_C}.
\end{equation}
Hence we find that for very large $N$ we may write, in effect:
\begin{equation}\label{eq:B.14}
\frac{d{\bar E}}{dT}\vert_{T=T_C} = \frac{(\ln 2)^2}{6}N +
\mathcal{O}(1),
\end{equation}
which is Eq. (5.2). $\mathcal{O}(1)$ denotes the terms, which are of
the order $N^0$, or less.


\begin{thebibliography}{40}
\bibitem{yksitoista} It is interesting that, up to an unimportant
numerical factor $2\ln 2$, this expression for energy is the same as
the one used as a starting point in a sketch for an entropic theory
of gravity by Verlinde in~\cite{yksitoistaa}.
\end{thebibliography}
\end{document}